\documentclass[aps,prb,reprint,twocolumn,showpacs,superscriptaddress,amsmath,amssymb]{revtex4-1}

\usepackage[utf8]{inputenc}
\usepackage{graphicx}
\DeclareGraphicsExtensions{.eps}
\usepackage{lipsum}
\usepackage{bm}
\usepackage{xparse}
\usepackage{textcomp} 
\usepackage{enumitem}
\usepackage{amsmath,amssymb}
	\usepackage{graphicx}
	\usepackage{color}
	\usepackage{xcolor}
	\usepackage{longtable}
	\usepackage{footnote}
	\usepackage{physics}
	
	\usepackage{dcolumn}
	\usepackage{array}
	\usepackage{blindtext, rotating}
	\usepackage{braket}
	\usepackage{multirow}
	\setlist{noitemsep,leftmargin=*}

	\makeatletter
	\usepackage{hyperref}
	\hypersetup{
		colorlinks=true,
		urlcolor= blue,
		citecolor=blue,
		linkcolor= blue,
		bookmarksopen=false,
	}

\begin{document}

\title{Influence of Interlayer Stacking on Optical  Behavior in  WSe$_{2}$/MoS$_{2}$  van der Waals Heterostructures}
	\author{Widad Louafi}
	\affiliation{Laboratoire de Physique Th\'eorique, Facult\'e des Sciences Exactes, Universit\'e de Bejaia, 06000 Bejaia, Alg\'erie}
	
	\author{Karim Rezouali}
	\affiliation{Laboratoire de Physique Th\'eorique, Facult\'e des Sciences Exactes, Universit\'e de Bejaia, 06000 Bejaia, Alg\'erie}
	\affiliation{Faculty of Physics and CENIDE, University of Duisburg-Essen, D-47053 Duisburg, Germany}
	\affiliation{Peter Grünberg Institut, Forschungszentrum Jülich and JARA, D-52425 Jülich, Germany}
	
	\author{Daniele Varsano}
	\affiliation{Centro S3, Istituto Nanoscienze—Consiglio Nazionale delle Ricerche (CNR-NANO), via Campi 213/A, 41125 Modena, Italy}
	\affiliation{European Theoretical Spectroscopy Facility (ETSF)}
	
	\author{Maurizia Palummo}
	\affiliation{Dipartimento di Fisica, Università di Roma ‘Tor Vergata’, Via della Ricerca Scientifica 1, I-00133 Roma, Italy}
	\affiliation{European Theoretical Spectroscopy Facility (ETSF)}
		
	\author{Maurits W. Haverkort}
	\affiliation{Institute for Theoretical Physics, Heidelberg University, Heidelberg, Germany}
	
	\author{Samir Lounis}
	\affiliation{Institute of Physics, Martin-Luther-University Halle-Wittenberg, 06099 Halle (Saale), Germany}
	\affiliation{Peter Grünberg Institut, Forschungszentrum Jülich and JARA, D-52425 
		Jülich, Germany}

\begin{abstract}
	We investigate the impact of crystal alignment on excitonic behavior in WSe$_{2}$/MoS$_{2}$ van der Waals heterostructures by comparing eclipsed (AA) and staggered (AB) stacking configurations. Our first-principles and symmetry-based analysis reveal that interlayer stacking symmetry plays a central role in determining the nature of electron-hole pairs. We uncover a rich variety of excitonic states, including spatially confined two-dimensional (2D) excitons, delocalized three-dimensional (3D) excitons, and charge-transfer (CT) excitons  with interlayer character. The dimensionality and optical activity of these excitons are governed by the interplay among orbital character, interlayer hybridization, and symmetry-imposed selection rules. Our findings establish general principles for engineering excitonic properties in van der Waals heterostructures through controlled layer orientation and stacking order.
\end{abstract}

\maketitle

\section{ Introduction }

Two-dimensional (2D) semiconducting materials made of transition metal dichalcogenides (TMDs) have recently gained popularity due to their intriguing physical and chemical properties, which hold promise for the next generation of optoelectronic, photonic, and energy conversion devices.\cite{Novoselov2016,Xia2014,Fang2012,Mak2010,Mak2016,Wang2012}

Among others, monolayers WSe$_{2}$ and MoS$_{2}$ are two important members of these materials, composed of a hexagonal sheet of transition metal atoms (Mo, W) sandwiched between two sheets of chalcogen atoms (S, Se), which are held together by covalent bonds. The category of these two 2D materials fabricated by means of mechanical exfoliation, are finite band gap materials exhibiting direct fundamental gaps spanning the visible to near-infrared (NIR) spectrum. This characteristics enables efficient light-matter interaction \cite{Splendiani2010} and opens a large avenue for applications in photodetectors, \cite{Lopez-Sanchez2013,Koppens2014} light-emitting diodes (LEDs),\cite{Ross2014,Pospischil2014,Baugher2014} and photovoltaic devices.\cite{Bernardi2013,Furchi2014} The strong excitonic effects observed in these materials further enhance their optical performance, making them ideal candidates for nanoscale optoelectronic devices.\cite{Glazov2018}

A critical trait of TMDs monolayers is that their electronic and optical properties are valley-dependent, a consequence of  strong spin-orbit coupling and broken inversion symmetry.\cite{Cao2012,Xu2014} The development of valleytronics, an emerging discipline that exploits the degree of freedom of valleys for information processing and storage, \cite{Di2012,Mak2014,Kormanyos2015} is made possible by the presence of distinct valleys at the $K$ and $K'$ points of the Brillouin zone. 

The electronic properties of TMDs monolayers can be engineered by creating heterostructures through stacking different monolayers. These heterostructures are kept together by weak van der Waals (vdW) forces, allowing  for precise control over layer composition, stacking orientation, and interlayer coupling. \cite{Geim2013,Novoselov2016,Chendong2017} A particularly intriguing configuration is the type-II band alignment, where the conduction band minimum (CB) of one material aligns with the valence band maximum (VB) of the adjacent layer. This staggered energy level arrangement enables the spatial separation of photoexcited charge carriers, making type-II heterojunctions worth using in photovoltaics, photodetectors, and photocatalysis. \cite{Lee2014,Yiling2016}

TMD heterostructures also exhibit efficient exciton dissociation and charge transfer processes, which are crucial to the performance of optoelectronic devices. These structures exploit the tunability of interlayer excitons, in which electrons and holes reside in distinct layers, resulting in prolonged excitonic radiative lifetime and consequently higher device efficiency. \cite{Palummo2015,Molina2013}   For instance, WSe$_{2}$/MoS$_{2}$ and WS$_{2}$/MoSe$_{2}$ heterostructures have shown extraordinary performance in charge carrier separation and energy conversion applications. \cite{Furchi2014,Cong2014} \cite{Furchi2014,Cong2014} Due to the presence of type-II band alignment and  tailored vdW interactions, these systems  were able to produce photocurrent and harvest energy more effectively. \cite{Britnell2013,Aggoune2017}

Going beyond optoelectronic applications, TMD heterostructures are emerging as promising platforms for quantum technologies. Twisting adjacent TMD layers to form moiré superlattices enables the emergence of flat electronic bands, which host correlated phenomena such as Mott insulating states and superconductivity. \cite{Regan2020,Tang2020}  The Moiré potential landscape that results offers a highly tunable platform for the investigation of strong electronic correlations and quantum phase transitions, thereby emphasizing their potential to advance both fundamental quantum science and  device engineering. Moreover, the integration of TMD heterostructures with other 2D materials, such as graphene and hexagonal boron nitride (hBN), broadens their portfolios for  device layouts. Graphene, with its high electrical conductivity and flexibility, functions as an ideal electrode material, while hBN provides an atomically smooth insulating layer with minimal charge traps. \cite{Britnell2012}  These hybrid systems enabled the building of innovative devices, such as tunneling transistors, vertical heterojunction diodes, and flexible electronics.\cite{Withers2015}

In this work, we investigate the optoelectronic properties of  WSe$_{2}$/MoS$_{2}$ bilayer van der Waals (vdW) periodic heterojunctions, focusing on the three stackings that correspond to local energy minima in R-type Moiré patterns.\cite{Chendong2017} Using a highly precise, state-of-the-art ab initio many-body approach, we analyze the spectra in detail and demonstrate that the nature and dimensionality of excitons in these heterojunctions can be systematically controlled. Our findings reveal the significant impact of stacking configurations on the optical excitations at the absorption onset. By systematically varying the arrangement of the WSe$_{2}$ and MoS$_{2}$ layers, we identify three distinct types of electron-hole pairs of two-dimensional (2D) intralayer excitons, three-dimensional (3D) interlayer excitons, and charge-transfer (CT) excitons.
We further discuss their emergence based on symmetry considerations, providing a deeper understanding of the relationship between stacking order and excitonic properties in vdW heterostructures.

The paper is organized as follows: Section II describes the computational methods employed in this study. Section III presents and discusses our results, and Section IV summarizes the main conclusions.

\section{Method }
We performed density functional theory (DFT),\cite{Hohenberg1964,Kohn1965} $GW$, and  Bethe–Salpeter equation (BSE) formalism utilizing the Quantum Espresso \cite{Giannozzi2009} and Yambo \cite{Marini2009, Sangalli2019} codes. This allows for a sufficiently accurate description of both the ground-state (DFT), one-particle excitation energies $(GW)$, and two-particle excitonic excitations as observed in optical experiments (BSE).

We first calculated the ground-state properties  utilizing DFT, \cite{Hohenberg1964,Kohn1965} with  norm-conserving pseudopotentials and a plane-wave basis set. The ionic potential was modeled using Optimized Norm-Conserving Vanderbilt (ONCV) pseudopotentials.\cite{Hamann2013}
The exchange-correlation effects were treated within the generalized gradient approximation (GGA) using the Perdew-Burke-Ernzerhof (PBE) functional, \cite{Perdew1996} supplemented by the vdW-D3 correction \cite{Grimme2010} to account for long-range van der Waals interactions between monolayers. Spin-orbit coupling was included in all calculations, and the $sp$ semicore states of the transition metal atoms were explicitly treated as valence electrons.\cite{Bernardi2013,Molina2013,Cheiwchanchamnangij2012} This approach is essential for accurately capturing the exchange contribution to the self-energy term in $GW$ calculations.

For all systems, we adopted a plane-wave energy cutoff of 70 Rydberg (Ry) for the electronic wave function expansion and employed a 12 $\times$ 12 $\times$ 1 Monkhorst-Pack \cite{Monkhorst1976} $k$-point mesh for Brillouin zone (BZ) sampling. A vacuum layer of 27 \AA ~  was introduced along the out-of-plane direction to prevent spurious interactions between periodic images.

Structural relaxations were performed until the atomic forces were reduced below 0.01 eV/\AA, and the energy convergence threshold was set to 10$^{-5}$ eV. These criteria ensured that the residual stress remained below 0.01 kbar.

Since DFT, in its local and semi-local approximations, often underestimates band gaps due to its limited treatment of electron-electron interactions, we employed the more accurate GW approximation to obtain reliable quasiparticle energies. Specifically, the $G_0W_0$ approach was used to correct the Kohn-Sham eigenvalues by solving the Dyson equation,\cite{Hedin1965, Hybertsen1985,Rohlfing2000,Onida2002} thereby renormalizing the band gaps to values in better agreement with experimental measurements. 
The quasiparticle (QP) energies (single-particle excitation energies) were computed within the $G_0W_0$ by using the linearized quasiparticle equation:\cite{guandalini2023}
\begin{equation}
\varepsilon_{\mathbf{QP}}^{n\mathbf{k}} = \varepsilon_{n\mathbf{k}} + Z_{n\mathbf{k}} \langle n\mathbf{k} | \Sigma(\varepsilon_{n\mathbf{k}}) - v_{xc} | n\mathbf{k} \rangle \;,
\end{equation}
where $\epsilon_{n\mathbf{k}}$, $\phi_{n\mathbf{k}}^{KS}$ and $V_{xc}$ describe the Kohn-Sham eigenvalues, orbitals, and  exchange and correlation potential over the n$\mathbf{k} $ KS eigenvectors. The normalization factor $Z_{n\mathbf{k}}$ is defined as: 
\begin{equation}
	Z_{n\mathbf{k}}= {\left[1-\left\langle n{\bf k}\middle|\frac{\partial\Sigma(\omega)}{\partial\omega} \middle| n{\bf k} \right\rangle \Bigg|_{\,\omega=\epsilon_{n{\bf k}}} \right]}^{-1} 
\end{equation}
The expectation values of the self-energy are written as:
\begin{equation}
\Sigma_{nk}=-\int \dfrac{d\omega'}{2\pi i} e^{i\omega' 0^{+}}\big\langle\phi_{n\mathbf{k}}^{KS}\big|G(\omega+\omega')W(\omega')\big|\phi_{n\mathbf{k}}^{KS}\big\rangle\;.
\end{equation}

The Green functions, $G$, are constructed using the DFT eigenfunctions, $\phi_{n\mathbf{k}}^{KS}$, and their corresponding eigenvalues, $\epsilon_{n\mathbf{k}}$. The dynamical screening effects in the self-energy, $W$, are treated within the generalized plasmon-pole model.\cite{Godby1989} In the GW self-energy calculations, cutoff energies of 70 Ry and 12 Ry are applied for the exchange and correlation components, respectively, and up to 350 empty states are included in the summation over unoccupied states. A uniform $k$-point grid of  27 $\times$ 27 $\times$ 1  is utilized for both monolayers and bilayer heterostructures to ensure accurate BZ sampling. BZ sampling. To avoid numerical divergence of the coulomb potential at small $\mathbf{q}$ which creates convergence problems in the  QP corrections of 2D semiconductors, we used the  so-called Random Integration Method.\cite{guandalini2023}

Starting from the Kohn-Sham wave functions and the quasiparticle energies, the absorption spectra are calculated on the level of Bethe–Salpeter equation (BSE)  formalism:\cite{Hanke1982,Rohlfing2000,Onida2002,Strinati1988}
\begin{equation}
\big(E_{c\mathbf{k}}-E_{v\mathbf{k}}\big)A^{S}_{vc\mathbf{k}}+\sum_{\mathbf{k'}v'c'}\big\langle{vc\mathbf{k}}\big|K_{eh}\big|v'c'\mathbf{k'}\big\rangle A^{S}_{v'c'\mathbf{k'}}=\Omega^{s}A^{S}_{vc\mathbf{k}}
\end{equation}
where $A^{S}_{vc\mathbf{k}}$, $\Omega^{s}$, and $K_{eh}$   are  the
expansion coefficients of the excitons in  electron-hole basis, the  eigenenergies, and the kernel of the BSE, respectively .

Here, the excitonic Hamiltonian is constructed in the basis of electron-hole pairs, accounting for direct excitations at a given momentum $\mathbf{k}$ from a valence band state with quasiparticle energy $E_v(\mathbf{k})$ to a conduction band state with quasiparticle energy $E_c(\mathbf{k})$. The interaction kernel, $K_{eh}$, incorporates both the repulsive unscreened exchange interaction ($V$) and the attractive screened Coulomb interaction ($W$) between electron-hole states.These interactions can give rise to discrete excitonic states below the onset of the quasiparticle continuum. 
In the absence of electron-hole interactions i.e., when the interaction kernel $K_{eh}$ is neglected, the optical excitations reduce to independent electron-hole pair transitions. In this case, the  optical response of the system  is governed by single-particle interband transitions without excitonic effects. 

The optical absorption spectrum is given by the imaginary part of the dielectric function, $\epsilon(\omega)$, and can be calculated as
\begin{align}
	\epsilon(\omega) =1 - \sum_{\mathbf{k}vc} \sum_{\mathbf{k'}v'c'} &\lim_{\mathbf{q}\rightarrow 0} \dfrac{8\pi}{|\mathbf{q}|^2 \Omega N_q}  \rho^{*}_{vc\mathbf{k}} (\mathbf{q},\mathbf{G})  \\ \nonumber
	&\times \rho_{v'c'\mathbf{k}'} (\mathbf{q'},\mathbf{G'}) \sum_{\lambda} \dfrac{A^{\lambda}_{vc\mathbf{k}} (A^{\lambda}_{v'c'\mathbf{k}'})^{*}}{\omega-E_{\lambda}}
\end{align}
 where $ \rho_{nm}$ are defined as \cite{guandalini2023}:
\begin{equation}
\rho_{nm}(\mathbf{k}, \mathbf{q}, \mathbf{G}) = \langle n\mathbf{k} | e^{i(\mathbf{q} + \mathbf{G}) \cdot \mathbf{r}} | m\mathbf{k + q} \rangle\;.
\end{equation}

The optical transitions are characterized by the dipole matrix elements, $\big\langle{c\mathbf{k}}\big|p_{i}\big|v\mathbf{k}\big\rangle$ , which describe the coupling between valence and conduction states. The excitonic states are defined by their excitation energies, $\Omega^{s}$, and corresponding amplitudes, $A^{S}_{vck}$. To achieve convergence in the solution of the Bethe-Salpeter equation (BSE), calculations were performed on the same $k$-point grid used in the GW  calculations. The interaction kernel in the BSE  was constructed using six conduction bands and six valence bands. A Lorentzian broadening of 0.15 eV was applied to simulate the absorption spectra. In both GW and BSE calculations, the Coulomb interaction was truncated to eliminate spurious interactions between periodic images of the layers.\cite{Rozzi2006}
 
\section{RESULTS AND DISCUSSION }
\subsection{Structural properties}

WSe$_2$ and MoS$_2$ monolayers are transition-metal dichalcogenides (TMDs) made up of two Se (S) layers and hexagonal planes of W (Mo) atoms covalently bound to them in a trigonal prismatic coordination. Through vertical stacking of these two-dimensional crystals, a vdW structure is created, forming the heterobilayer WSe$_2$/MoS$_2$. Despite rotational alignment during fabrication, the small in-plane lattice mismatch (3-4 \%) between WSe$_2$ and MoS$_2$, generates Moir{\'e} patterns.\cite{Bernal1924,Constantinescu2013,Chendong2017}

\begin{figure*}[t]
	\begin{center}  
	\includegraphics[scale=0.15]{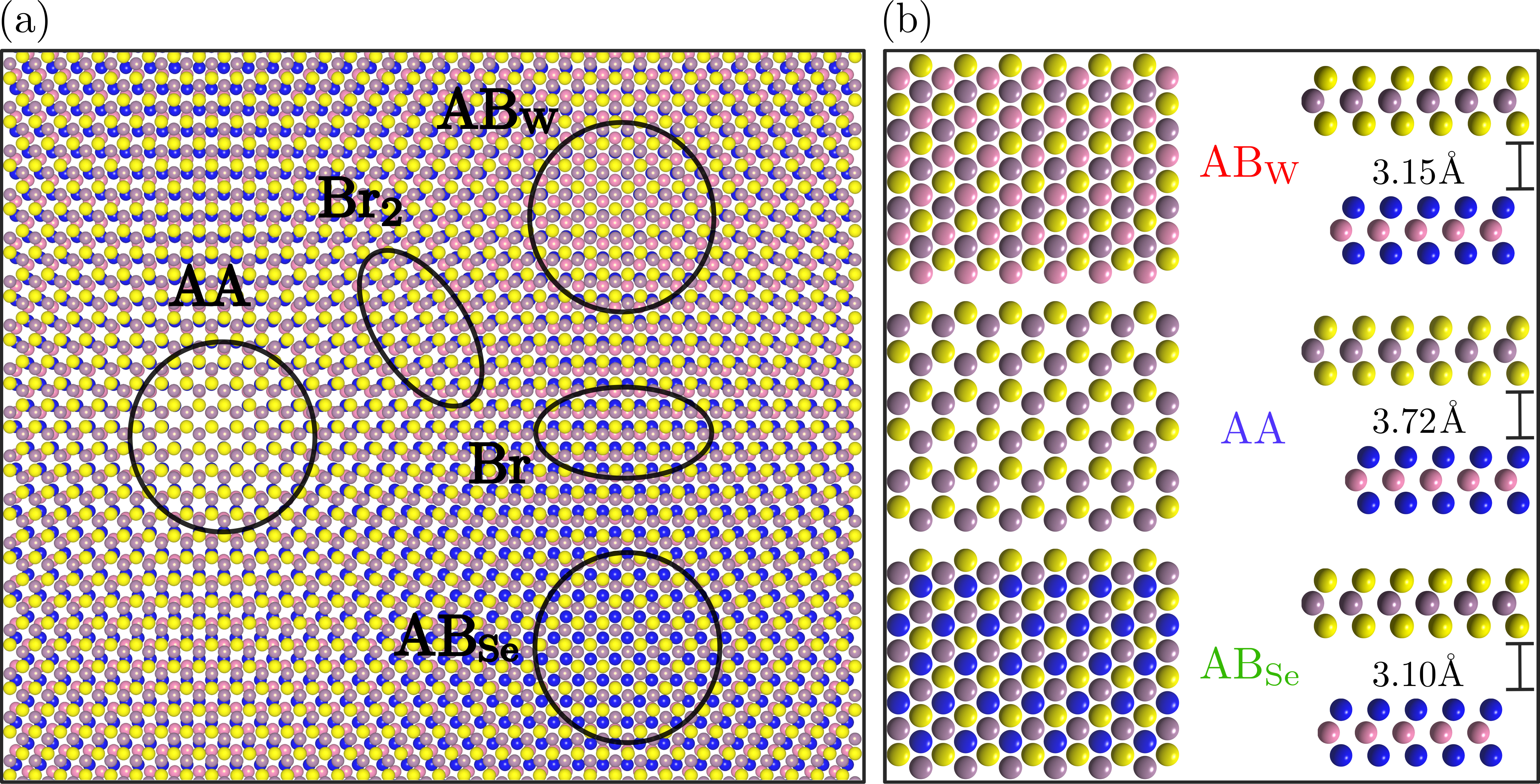}
	\end{center} 
	\caption{(a) Schematic model of the Moiré pattern
		on an R-stacked  WSe$_{2}$/MoS$_{2}$ heterobilayer. (b) Top and side views  of the atomic models for AA, AB$_{\text{W}}$, and AB$_{\text{Se}}$ registries zoomed in on a unit cell of the Moiré pattern.}
	\label{Fig1}
\end{figure*}

An atomic model of the R-type Moiré supercell is depicted in Figure \ref{Fig1}(a), which follows the nomenclature of Ref. \cite{Chendong2017} (itself adapted from conventional stacking terminology \cite{Bernal1924,Constantinescu2013}) and highlights three local energy-minimum configurations labeled AA, AB$_{\text{W}}$, and AB$_{\text{Se}}$. In  Fig. \ref{Fig1}(b), we  show the atomic arrangements for these three local alignments. These arrangements correspond to the three stable energy-minimized states found in the heterobilayer, and the interaction between lattice mismatch and interlayer registry is reflected in the Moiré supercell geometry. Two local energy-maximum bridge configurations (Br and Br$_2$) are also accessible. These intermediate structures are produced through translation of the MoS$_{2}$ lattice with respect to WSe$_{2}$ between neighboring AB$_{\text{Se}}$ sites. The arrangement of local atomic structures within these stacking configurations leads to a modification of interlayer forces, resulting in out-of-plane corrugations and spatially dependent  distances between layers within the heterostructure. \cite{Bernal1924,Constantinescu2013,Chendong2017}

To model each labeled site in Fig.~\ref{Fig1}, a $1\times1$    WSe$_{2}$/MoS$_{2}$ heterobilayer unit cell was constructed, containing six atoms. The lattice constant was set to the average of the monolayer WSe$_{2}$ and MoS$_{2}$ lattice parameters, preserving the hexagonal symmetry inherent to the parent monolayers. The three stable stacking configurations examined here (AA, AB$_{\text{W}}$, and AB$_{\text{Se}}$) were predicted theoretically,\cite{Chendong2017} and  observed experimentally in samples grown by chemical vapor deposition (CVD).~\cite{Chendong2017,Lin2015}
The AA stacking is constructed by positioning Mo atoms over W atoms and S atoms in one layer over Se atoms in the layer beneath with zero degree rotation of WSe$_{2}$ with respect to the MoS$_{2}$ layer. The difference in the local atomic registry between AA and AB stackings resides in the lateral alignment of the chalcogen atoms. AB$_{\text{Se}}$ (AB$_{\text{W}}$) means that the hexagonal lattices of the metal atoms in the two monolayers are stacked in an AB manner with the Mo (W) atoms in the one layer are  covered by chalcogen atoms in the other layer. In other words,  W atoms is covered by S atoms in AB$_{\text{Se}}$ and   Mo atom is covered by Se atoms in AB$_{\text{W}}$. Consequently, interlayer coupling would be significantly  influenced as we will see later in the paper.

According to the nomenclature of the previous  study, \cite{Constantinescu2013} AA is an eclipsed stacking. The AB bernal stacking can be obtained by simple transformations from AA configuration by horizontal layer sliding. AB is staggered with metal atoms of one layer aligned with chalcogen atoms of the other layer and can be obtained by shifting the WSe$_{2}$ layer with respect to MoS$_{2}$ layers by $a/\sqrt{3}$, where $a$ is the lattice parameter. Both AA and AB stacking orders are of R-type. 

We computed the lattice constants, interlayer distances, and  bond lengths  for monolayers and bilayers. The lattice constant of the heterobilayer is approximately 3.23 \AA, achieved through a 2.7~\% in-plane compression of the monolayer WSe$_{2}$ and a 1.5~\% expansion of the monolayer MoS$_{2}$ to enforce commensurability. In the three stacking configurations, the average Mo-S and W-Se bond lengths are 2.425 \AA ~and 2.528 \AA, respectively. These correspond to compressions of about 0.3~\% (Mo-S) and 0.8~\% (W-Se) relative to their isolated monolayer values. The minor deviations in lattice parameters and bond lengths from those of the parent monolayers are consistent with weak interlayer interactions. Our results align closely with prior theoretical and experimental studies.\cite{Yi2011,Chang2013,Ramasubramaniam2012,Chendong2017,Le2015}
\begin{figure*}[t]
	\center
	\includegraphics[scale=0.30]{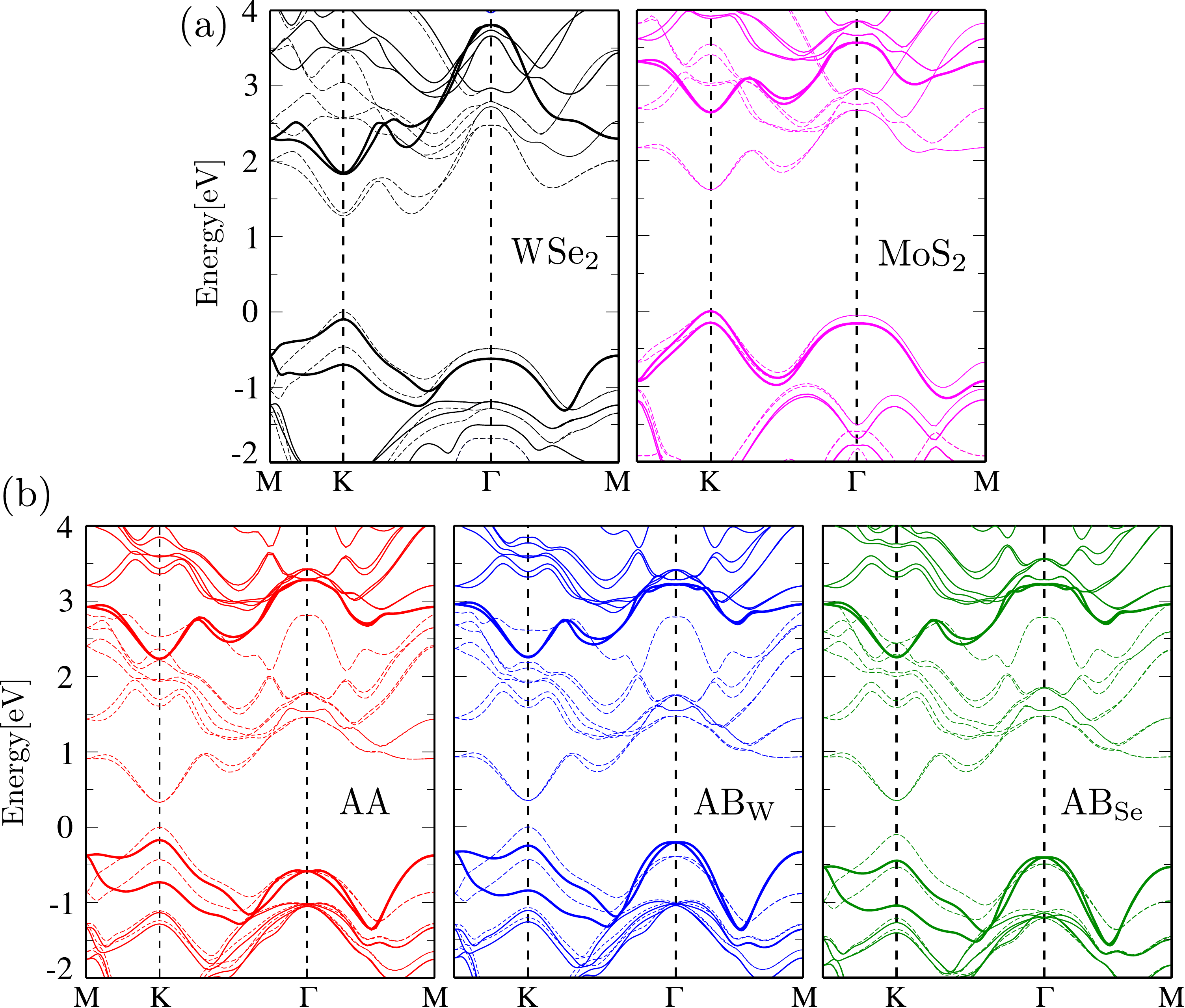}
	\caption{ (Color online) Band structures calculated DFT (GGA, thin dashed lines) and the $G_0W_0$ approximation (thick solid lines). Top panel: Monolayer MoS$_2$ (left) and monolayer WSe$_2$ (right). From left to right  in the bottom panel: AA, AB$_{\text{W}}$, and AB$_{\text{Se}}$ stacked  WSe$_2$/MoS$_2$ heterobilayers.}	
	\label{Fig2}
\end{figure*}

To evaluate the relative stability of the three configurations, we calculated their formation energies. Contributions from phonons were omitted due to the dominance of weak vdW interactions, \cite{Slotman} and zero-point vibrational effects were assumed negligible.\cite{Milko} The AA stacking configuration exhibits a formation energy of -223 meV/cell, which is only 0.06 meV and 0.07 meV lower than those of AB$_{\text{W}}$ and AB$_{\text{Se}}$, respectively. Although AA stacking is the most stable configuration, the very small energy differences between the three configurations indicate that they have comparable thermodynamic stability.

The interlayer distance $d$, which is the distance  between the constituent  layers [see Fig. \ref{Fig1}(c)], is calculated to be approximately 3.72 \AA, 3.15 \AA, and 3.10 \AA~ for AA, AB$_{\text{W}}$, and AB$_{\text{Se}}$, respectively. This  distance is largest when the S atoms are directly aligned above the Se atoms, as this is the case in the AA registry.   This occurs due to increased  Pauli repulsion between the overlapping electron clouds localized around these atoms. On the contrary, in AB stacking configurations, the interlayer distance $d$ is reduced due to the alignment of the metal atoms in one layer with the chalcogen atoms in the adjacent layer. This alignment strengthens the electrostatic attraction, resulting in a reduced separation between the Se and S atomic layers.

\subsection{Electronic properties }
Figure \ref{Fig2} shows the DFT and G$_0$W$_0$ band structures for monolayer WSe$_{2}$ and MoS$_{2}$, and their corresponding heterostructures. Consistent with prior studies,\cite{Cheiwchanchamnangij2012,Molina2013,Ramasubramaniam2012,Dadkhah2024,Qiu2013} both WSe$_{2}$ and MoS$_{2}$ monolayers exhibit direct band gap semiconducting behavior at their equilibrium geometries under  DFT and G$_0$W$_0$  treatment [see Fig.\ref{Fig2} (a)], a direct consequence of inversion symmetry breaking (group of symmetry $D_{3h}$) inherent to their honeycomb lattice structures. \cite{Mak2010,Splendiani2010}
Our mean field theory calculations  reveal fundamental gaps of 1.62 eV (MoS$_{2}$) and 1.27 eV (WSe$_{2}$) at $K$ point, with doubly degenerated valence bands  predominantly of localized $d_{x^2-y^2}$ character. Spin-orbit (SO) coupling induces a splitting of the valence band into two uppermost valence bands (VB and VB-1), with energy separations of 457 meV in  WSe$_{2}$ and 147 meV in MoS$_{2}$, consistent with prior theoretical and experimental reports.\cite{Cheiwchanchamnangij2012,Diana2013,Ramasubramaniam2012,Le2015,Dange2025,Terrones2013,Shih2024} Many-body  G$_0$W$_0$ corrections substantially increase these fundamental gaps to 2.72 eV and 2.17 eV, respectively, showing excellent consistency with established theoretical references. \cite{Cheiwchanchamnangij2012,Kadantsev2012,Molina2013,Diana2013,Shih2024,Dange2025} Discrepancies with values available in the literature likely originate from variations in computational parameters, particularly exchange-correlation functional implementations and unit cell optimizations.

For heterobilayers [Fig. \ref{Fig2}(b)], DFT calculations reveal direct band gaps at the  $K$ point, with twofold degenerate valence band. The computed band gaps are 0.33 eV (AA), 0.35 eV (AB$_{\text{W}}$), and  0.45 eV (AB$_{\text{Se}}$), in agreement with existing literature.\cite{Terrones2013,Dange2025} Crucially, the bandgap tunability with stacking order suggests a pathway to tailor electronic properties via layer arrangement. Near the $K$-point, hybridization effects dominate the band edges, driven by the $d$ orbital contributions of the metal atoms: the valence band arises from hybridization of W 5$d_{x^2-y^2}$ and 5$d_{3z^2-r^2}$  orbitals, while the conduction band is predominantly derived from Mo 4$d_{3z^2-r^2}$ orbitals. To model the local configurations, we employed $1 \times 1$ WSe$_2$/MoS$_2$ unit cells with a fixed in-plane lattice constant of 3.23~\AA, corresponding to the average of the unstrained monolayer lattice parameters. In this configuration, the MoS$_2$ layer experiences tensile strain, while the WSe$_2$ layer is under compressive strain. Although the influence of this strain on the interlayer interactions is expected to be minor, we assessed its effect by comparing the electronic band structures of unstrained and strained monolayers. Our calculations indicate that the band gap of monolayer WSe$_2$ increases by approximately 3 \%, while the band gap of monolayer MoS$_2$ decreases by about 7 \%.
 
Interlayer vdW interactions in WSe$_{2}$/MoS$_{2}$ heterobilayers yield a type-II band alignment, wherein the valence band  and the conduction band are localized on distinct monolayers. This spatial separation of charge carriers generates momentum-direct charge-transfer excitons, with holes confined to WSe$_{2}$ and electrons to MoS$_{2}$. Such a mechanism profoundly influences optical properties and presents compelling opportunities for optoelectronic applications, including excitonic devices and light-harvesting systems.
\begin{figure*}[t]
	\center
	\includegraphics[scale=0.25]{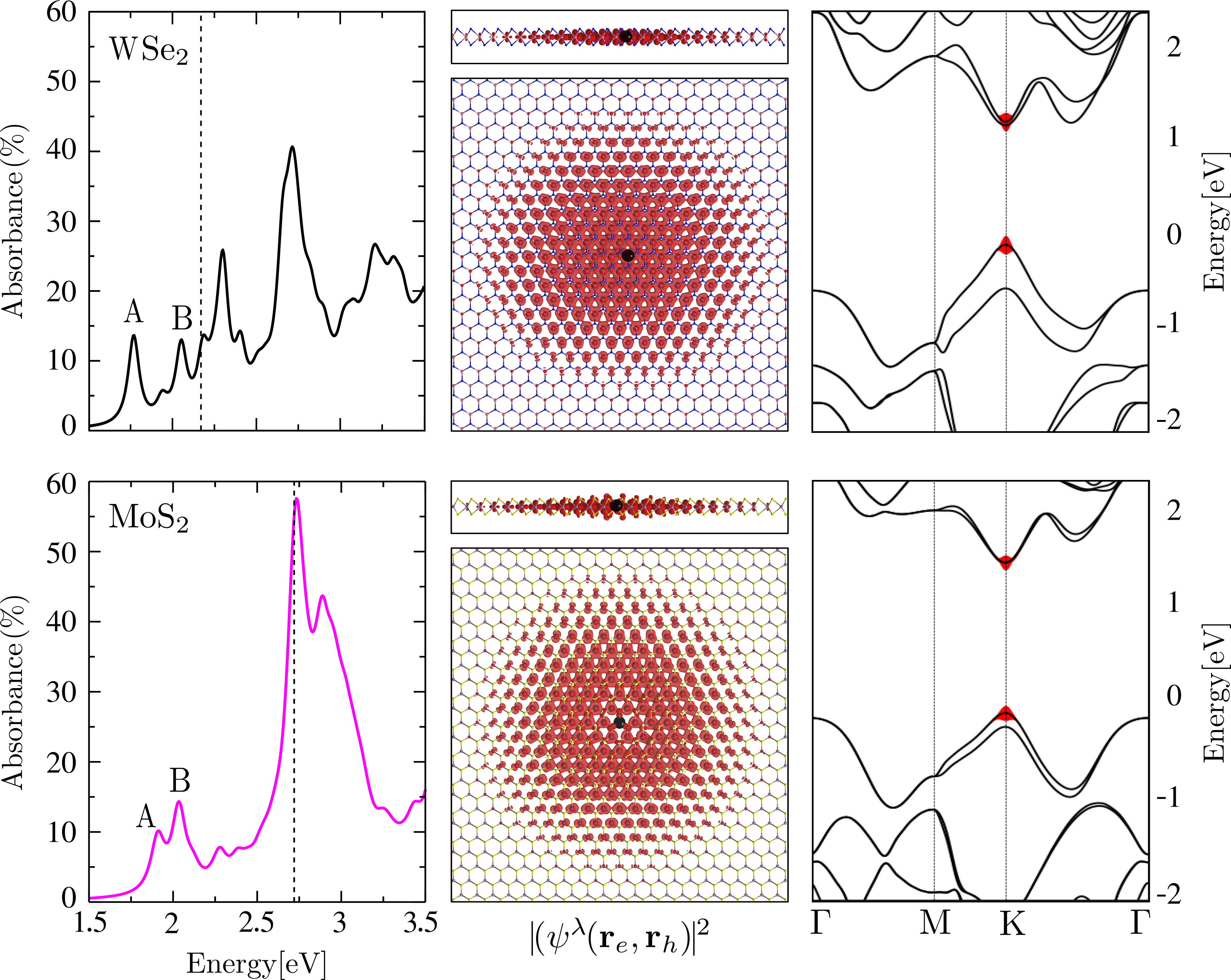}
	\caption{ (Color online) Left: Optical absorption spectra of  WSe$_{2}$  and MoS$_2$ single layers, given by the absorbance. QP band gaps are marked with vertical dashed lines, and the labels indicate the peaks associated with the A and B excitons.  Middle:  Side view  and top view  of exciton A charge density. Right: Reciprocal-space distribution of the  exciton A.  The sizes of the red patches  are indicative of the weights of the involved QP states.}
	\label{Fig3}
\end{figure*} 
At the DFT level, heterobilayer band gaps are typically smaller than those of the individual monolayers due to interlayer interactions and orbital hybridization between the two layers. Specifically, the $d_{3z^2-r^2}$ orbitals of W and Mo atoms overlap across the layers, modifying the band structure through interlayer coupling. These $d$ orbitals are more localized than $s$ and $p$ orbitals and often exhibit strong electron-electron repulsion, leading to huge electronic correlation effects.  Standard DFT, especially with local or semi-local functionals such as GGA, often struggles to capture these correlation effects accurately, particularly for the localized $d$ electrons. Consequently, DFT may not account for the strong Coulomb interactions within the $d$ orbitals, which reduces the energy difference between the valence 5$d$ bands and the   conduction 4$d$ bands, thereby leading to a smaller fundamental band gap in the bilayer system compared to the parent monolayers.

Moreover, accounting for electron-electron interactions through a G$_0W_0$ calculation significantly enhances the band gap by over 80\%. This result aligns closely with prior theoretical studies.\cite{Dange2025} Interestingly, the AA stacking configuration retains its direct band gap character at $K$, while AB stackings transition to indirect band gap materials upon the inclusion of electron-electron self-energy.
The origin of this difference is the interlayer interaction.  While the CB remains at the $K$ point, the VB shifts from $K$ to $\Gamma$. This shift can explained by the differences in the effective mass of the charge carriers at these hight-symmetry points. The electron-electron interactions modify the effective mass, which affects the energy dispersion near the band edges. In AB stacking configurations, the effective mass at the $\Gamma$ point would be lighter compared to $K$ point making it energetically favorable for the VB to relocate, resulting in an indirect band gap in the AB stacking configuration. The AA registry is predicted to exhibit a direct band gap of 2.40 eV, highlighting its potential for optoelectronic applications due to its suitability as an efficient light emitter. For AB$_{\text{W}}$ and  AB$_{\text{Se}}$ registries, direct band gaps of 2.50 eV and 2.70 eV are observed at the $K$-point, while indirect gaps of 2.44 eV and 2.65 eV are found between the $\Gamma$ and K points. The small difference between the direct and indirect band gaps suggests that these materials may also be promising for optoelectronic applications, as they remain effective light emitters. 
The large quasiparticle corrections to the PBE band gap originates from the enhancement of the  electron-electron interaction and the weakening of the screening response in two-dimensional materials. Such a low screening gives rise to bound excitons with large binding energies. 
\begin{figure*}[t]
	\center
	\includegraphics[scale=0.3]{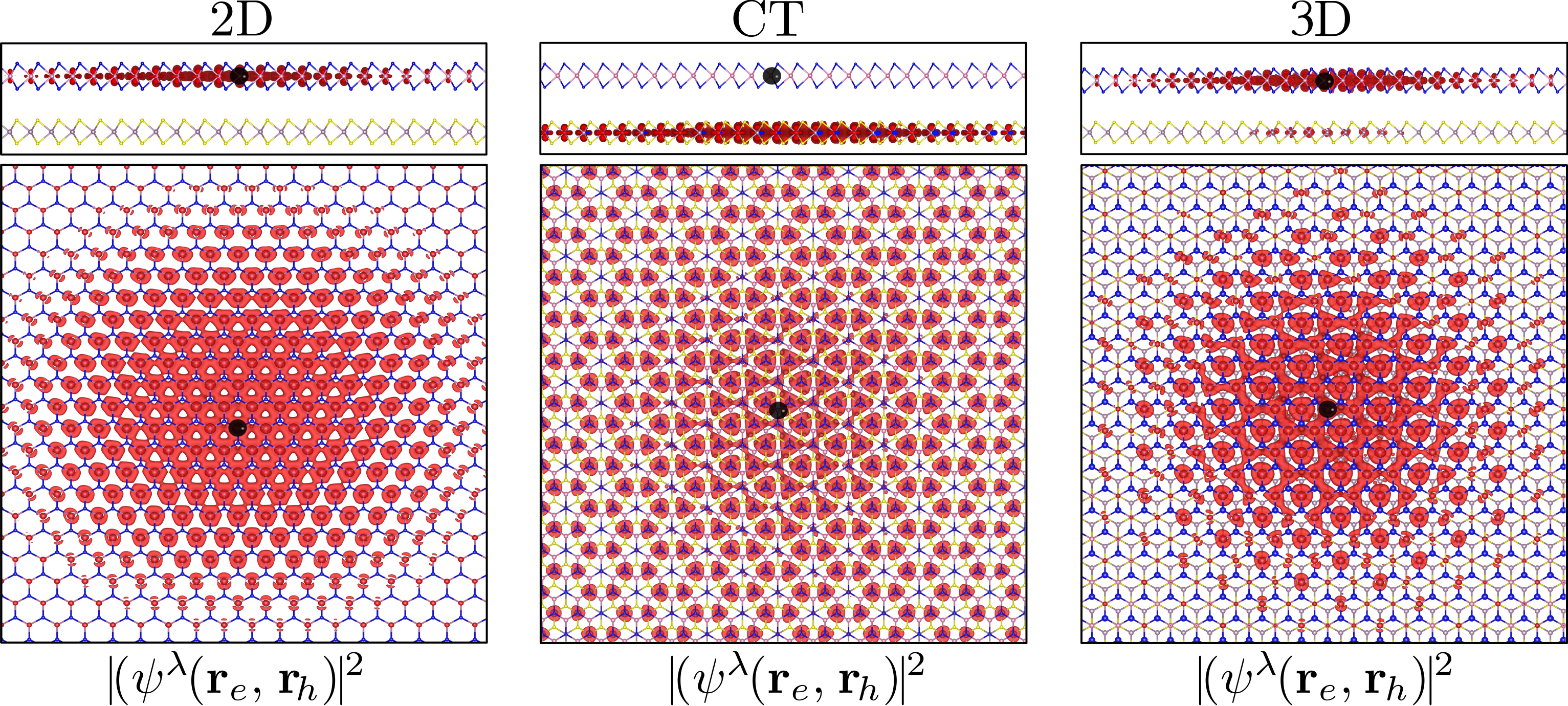}
	\caption{ (Color online) Side view  and top view  of
		exciton of charge densities (side and top view) of the three types of excitons that can emerge in WSe$_{2}$/MoS$_{2}$ heterobilayer upon different stacking configurations.
		From left to right: a 2D exciton, where the electron and hole reside within the same layer; a charge-transfer (CT) exciton, with the electron and hole pinned in different layers, and 3D exciton where  the hole is located in one layer while the electron is distributed across both layers. The hole is indicated by a black dot, while the electron distribution is depicted by the red isosurface.}
	\label{Fig4}
\end{figure*}
The bilayers possess the factor group $C_{3v}$, which also lacks inversion symmetry. This lack of symmetry, combined with SO coupling, result in valence band splitting. The values are 562 meV, 594 meV, and 600 meV  for AA, AB$_{\text{W}}$, and AB$_{\text{Se}}$, respectively, demonstrating a marked enhancement compared to monolayer counterparts. This increased splitting may arise from interlayer coupling, which introduces an additional contribution to the SO interaction beyond monolayer effects.
\subsection{Optical  absorption spectra  }
In figure \ref{Fig3} are shown the optical absorption spectra of monolayer  WSe$_{2}$ (top left panel) and MoS$_{2}$ (bottom left panel), revealing striking similarities in their low-energy excitonic features.
All systems exhibit a characteristic double-peak structure at the absorption onset (denoted as A and B excitons). These spectral features come from strong interactions between electrons that create tightly bound excitons, which mainly influence how the material responds to light.  The energy splitting between A and B excitons reflects spin-orbit coupling (SOC) effects in the valence band. Monolayer MoS$_{2}$ exhibits a splitting of 114 meV, while WSe$_{2}$ shows a larger splitting of 183 meV, attributed to the heavier tungsten atom enhancing SOC. 
The A (B) exciton  is found at 1.76 eV  (1.93 eV) and 1.91 eV (2.03 eV) in MoS$_{2}$ and WSe$_{2}$ monolayers, respectively, with binding energies of 0.81 eV (A) and 0.61 eV (B) for  MoS$_{2}$, and 0.40 eV (A) and 0.23 eV (B) for WSe$_{2}$, which  values significantly exceeding those in bulk semiconductors. Our results are in good agreement with previous calculations\cite{Marsili2021,Dadkhah2024} and experiments.\cite{Schmidt_2016} This enhancement stems from reduced dielectric screening in atomically thin layers, which amplifies Coulomb interactions.

The lowest-energy exciton is optically forbidden in these structures, despite broken inversion symmetry. While the presence or absence of inversion symmetry in a crystal fundamentally dictates the optical activity of excitonic states, additional factors such as spin-orbit coupling (SOC) and band-edge orbital symmetries determine whether the lowest-energy exciton is optically active (bright) or inactive (dark). This distinction becomes apparent when comparing the excitonic behavior of these monolayers with that of hexagonal boron nitride (hBN), \cite{Aggoune2018} despite both of which crystallize in the non-centrosymmetric D$_{3h}$ point group. In hBN, the weak SOC and $\pi \rightarrow \pi^*$ character of the band edges ensure that the lowest exciton remains bright, as the dipole-allowed transition is symmetry-permitted. In contrast, monolayer WSe$_{2}$ and MoS$_{2}$ exhibit strong SOC, which lifts the spin degeneracy of both valence and conduction bands, giving rise to a spin-forbidden exciton that can lie below the bright A exciton in energy. Despite the lack of inversion symmetry, which typically permits dipole transitions, the excitonic ground state in monolayer WSe$_{2}$ and MoS$_{2}$ can therefore be dark. These observations drive home that, while inversion symmetry sets the general optical selection rules, the fine structure of excitonic states and the relative ordering of bright and dark excitons depend sensitively on SOC strength, orbital hybridization, and symmetry of the band edges.

In the middle panel of Fig. \ref{Fig3}, we plot the squared magnitude of the real-space wave function for the exciton corresponding to peak A in monolayer WSe$_{2}$ and MoS$_{2}$, with the hole localized near a W atom in WSe$_{2}$ and a Mo atom in MoS$_{2}$. The spatial profile of this exciton exhibits a nearly circular symmetry, closely resembling the 1$s$ orbital of a 2D hydrogenic system. In both monolayers, the electron density for the exciton in both materials predominantly localizes on nearest-neighbor metallic atoms adjacent to the hole, reflecting restricted spatial charge separation. This pronounced overlap between electron and hole wave functions indicates shorter excitonic radiative lifetimes. The electron density profiles exhibit rapid spatial decay, characteristic of tightly bound electron-hole ($e-h$) pairs. Notably, the exciton MoS$_{2}$ displays a more localized charge distribution compared to its counterpart in WSe$_{2}$, consistent with its highest binding energy. This trend indicates a direct correlation between binding energy and spatial delocalization: reduced binding energy corresponds to progressively extended excitonic wave functions.

As illustrated in the right panel of Fig. \ref{Fig3}, this exciton originates from electronic transitions involving the in-plane $d_{z^2-r^2}$ orbital in the VB band and the out-of-plane $d_{z^2}$ orbital in the  CB  of the metal atoms. It is important to note that the real-space wave function of exciton B is spatially indistinguishable from that of exciton A. This equivalence arises from the conservation of spin as a good quantum number at the $K$ and $K'$ valleys.\cite{Di2012} Despite spin-orbit-induced splitting in the valence bands, the spatial components of the valence band wave functions remain identical across both valleys. Consequently, peaks A and B, which stem from transitions between the spin-orbit-split valence bands (VB, VB-1) and the lowest conduction band (CB) at the $K$ and $K'$ valleys, respectively, share identical spatial wave functions. Although these excitons originate from distinct valleys and exhibit different spin configurations, their spatial structures are invariant under spin conservation, reflecting the symmetry imposed by valley-dependent spin selection rules.

Beyond the A and B peaks, the absorption spectrum contains contributions from weakly bound electron-hole pairs. At higher energies, the absorption increases abruptly, being in accordance with direct band-to-band transitions, where excitonic effects become negligible. Subtle variations in peak shapes and spectrum breadth across materials arise from differences in carrier effective masses, SOC strength, and dielectric screening mechanisms. Our findings exhibit excellent agreement with prior experimental and theoretical studies, \cite{Shih2024,Molina2013,Palummo2015,Qiu2013} validating the robustness of the observed trends.
\begin{figure}[t]
	\center
	\includegraphics[scale=0.22 ]{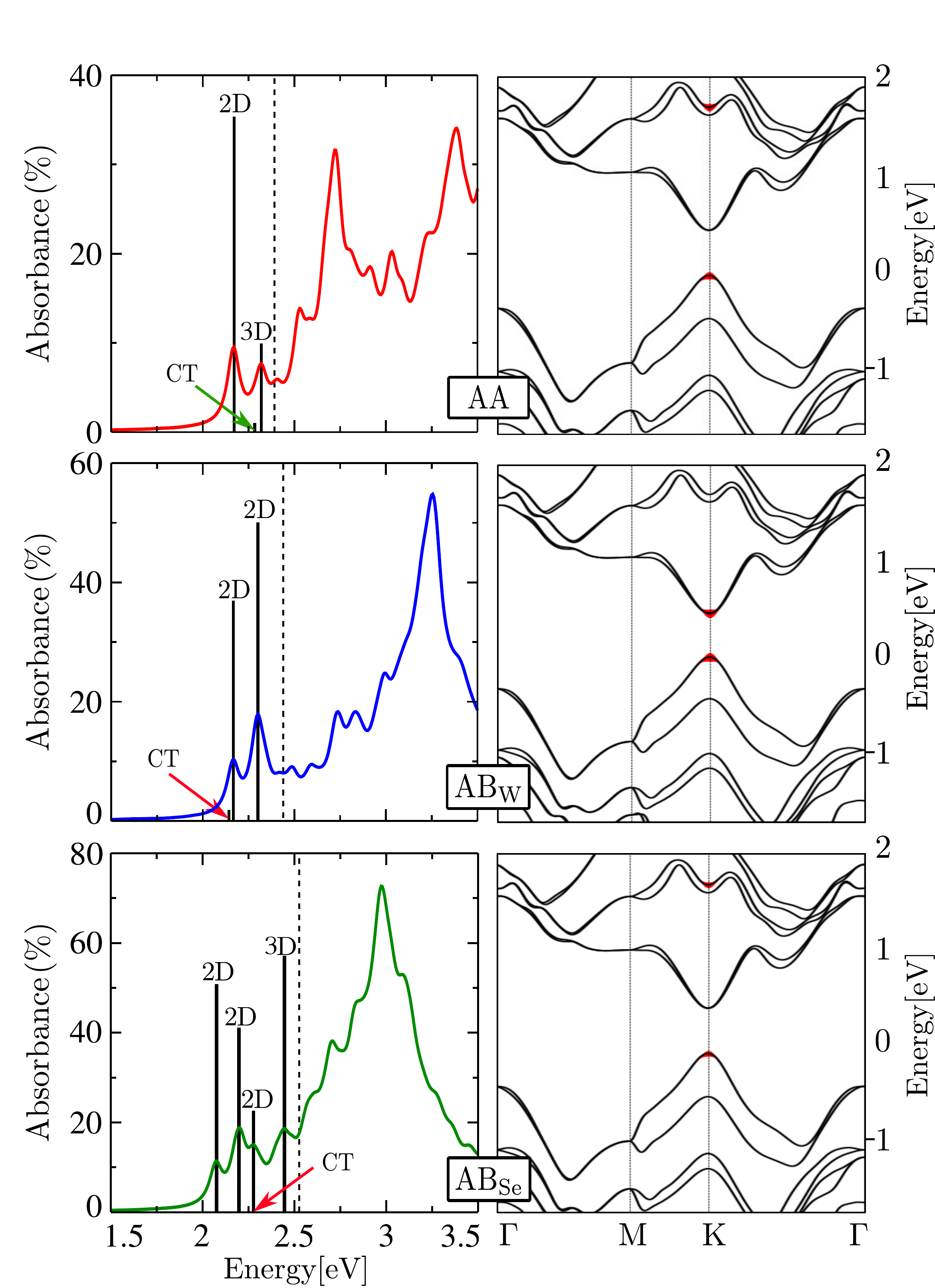}
	\caption{ (Color online) Left:  Optical absorption spectra, given by the absorbance in AA, AB$_{\text{W}}$ and AB$_{\text{Se}}$ registries. The dashed line indicates the direct QP gap. Right: Band (DFT) contributions to the excitations marked in the spectra following the nomenclature in Fig. \ref{Fig4}.  The sizes of the red patches  are indicative of the weights of the involved states. The 2D exciton of AA  stacking exhibits WSe$_2$-like character, whereas those of the AB stacking are of MoS$_2$-like.}
	\label{Fig5}
\end{figure}

We now turn to the excitonic properties of heterobilayers. In Fig. \ref{Fig4}, we display the real-space distributions of three exciton types found in different stacking configuration. The first type is a 2D exciton, with the electron and the hole settle in the same layer. These excitons can be further classified into MoS$_{2}$-like exciton (2D-MoS$_{2}$) and WSe$_{2}$-like exciton (2D-WSe$_{2}$) states in the in-plane direction. The second type is the charge transfer (CT) excitons, where the electron and the hole sit on different layers. The third category is the three-dimensional (3D) exciton, where the hole is confined to one layer while the electron is distributed across both layers.

In all cases, the excitonic wave functions exhibit trigonal symmetry, which mirrors the hexagonal lattice symmetry of the monolayer. Moreover, their charge distributions display extended features, indicative of localization in reciprocal space [see Fig. \ref{Fig5}]. The stacking-dependent structural and electronic properties are reflected in different optical responses. Specifically, the AA and AB${_\text{Se}}$ stacking arrangements accommodate the three exciton types, whereas the AB${_\text{W}}$ configuration supports only 2D and CT excitons. These excitonic characteristics are intimately linked to the symmetry and atomic registry of each stacking arrangement. To explain these features, we analyze the electronic band structures and optical spectra, emphasizing their interplay with the underlying atomic geometry.

In the eclipsed AA stacking, all atoms in the two nonequivalent layers align perfectly, with Mo atoms positioned directly above W atoms and S atoms directly above Se atoms. In contrast, the AB staggered stacking adopts a hollow-site registry, wherein alternating atoms occupy hollow sites. Specifically, in the AB${_\text{Se}}$ stacking, Mo and Se atoms are located at hollow positions, whereas in AB${_\text{W}}$, W and S atoms occupy these sites [see Fig. \ref{Fig1}(b)]. These different lateral registries result in varying electronic structures [see Fig. \ref{Fig2}(b)], which in turn influence the optical spectra [see Fig. \ref{Fig5}]. In the following discussion, we examine how these structural properties govern the formation of various $e-h$ pairs.

In  Fig. \ref{Fig5} we display the optical spectra (left panel) together
with the band structures (right panel) [see also Fig. S1 in the Supplemental Material \cite{Supporting}]. Our calculations reveal a significant redistribution of oscillator strength accompanied by a pronounced redshift of the absorption spectrum. Notably, we observe the emergence of strong excitonic resonances below the single-particle continuum onset, along with a large manifold of optically forbidden excitonic states below the onset of single-particle transition continuum 

First, we begin by examining the 2D exciton depicted in Fig. \ref{Fig4} [see also  Fig. S2 left panel in the Supplemental Material \cite{Supporting}]. This bound electron-hole ($e-h$) pair is observed across all three stacking arrangements, manifesting as the first bright excitons in each registry. In the AA configuration, it is non-degenerate and emerges at an energy of 2.17 eV, with a binding energy of 0.23 meV. Similar to the A exciton in monolayers, its wave function envelope is nearly azimuthally symmetric. This exciton marked in the spectrum of Fig. \ref{Fig5} arises from transitions between the VB and the CB+3 states near the band extrema at the high-symmetry $K$ point, as shown in Fig. \ref{Fig5} (right panel). Since both the VB and the CB+3 originate from orbitals localized in the W-based monolayer, hole localization within this layer confines the electronic charge density to the same layer, confirming its 2D character (2D-WSe$_{2}$). In AB stacking configuration, this exciton is  on MoS$_2$ layer (2D-MoS$_{2}$), and  due to the symmetry of the lattice it exhibits a twofold degeneracy. In the AB$_{\text{Se}}$ configuration,  it emerges at 2.07 eV with a notably high binding energy of 0.58 eV, while in the AB$_{\text{W}}$ configuration, it appears at 2.16 eV with a binding energy of 0.27 eV. In both cases, these excitons arise from transitions between the VB-2 and CB/CB+1 states at the $K$ point. Notice that the electron density associated with the 2D exciton is primarily localized on the nearest-neighbor metal atoms adjacent to the hole, resulting in restricted spatial charge separation and, consequently, reduced excitonic radiative lifetimes. 

Next, we examine the CT excitons, also depicted in Fig. \ref{Fig4}. These doubly degenerate optically forbidden excitons are also found in all stacking configurations, appearing as the first  exciton in all registries.  In AA,  the CT exciton  marked in the spectrum of Fig. \ref{Fig4} emerges at about  2.28 eV. It  is twofold degenerate, with a binding energy of 0.23 eV, and  originates from interband transitions between the valence and conduction bands at the high-symmetry $K$ point. In the spectrum of  AB${_\text{W}}$ configuration, the  CT exciton observed at 2.16 eV, has a binding energy of 0.28 eV, and arises from transitions between the VB and CB near $K$ point. In the AB${_\text{Se}}$ arrangement, the CT exciton  located at 2.27 eV  has a binding energy of 0.38 eV. It originates from transitions between the VB and the CB/CB+1, and between the VB and the CB, at the $K$ point. Here, the hole is localized on the  W$dz^2-y^2$ orbitals and the corresponding electronic wave functions are spread over metallic atoms belonging to  MoS$_2$ layer. Further details are provided  in the Supplemental Material. \cite{Supporting}

The final exciton type, 3D, is found exclusively in AA and AB${_\text{Se}}$ stacking configurations, appearing as the second bright exciton. It is a doubly degenerate exciton and appears at 2.32 eV with a binding energy of 0.08 eV, stemming from transitions between the VB and CB+4 states at the high-symmetry $K$ point. This wave function of this exciton is delocalized over the whole space and displaying extended features. In AB${_\text{Se}}$, the equivalent exciton appears at 2.44 eV with a binding energy of 0.21 eV. It is also degenerate and originates from transitions between the VB and CB+4 states in the vicinity of the $K$ point. Interestingly, this bright exciton demonstrates pronounced localization, extending over just a few lattice constants. Interestingly, we observe a characteristic dumbbell-like distribution of electron density around each W atom in all exciton types. This feature arises from the $d$ orbitals of individual metallic atoms and persists across all stacking configurations [see Fig. \ref{Fig5} and Fig. S2 in the Supplemental Material \cite{Supporting}].

The absence of 3D excitons in the AB${_\text{W}}$ stacking configuration is due to a complex interplay of symmetry constraints, interlayer hybridization effects, and orbital-specific interactions. In the AB${_\text{W}}$ registry, where W and S atoms occupy hollow sites, the resulting potential landscape and orbital overlap differ fundamentally from the AB${_\text{Se}}$ case. The 3D excitons observed in AA and AB${_\text{Se}}$ stackings originate from transitions to highly delocalized CB+4 states that require strong interlayer hybridization of Mo-d and W-d orbitals, a condition that is apparently not satisfied in AB${_\text{W}}$ due to the more localized nature of W-derived states near the Fermi level. This localization originates from both the stronger d-orbital confinement of W atoms and the particular nodal structure of the electronic wavefunctions in AB${_\text{W}}$, which restricts charge delocalization across the heterobilayer. Furthermore, the AB${_\text{W}}$ band structure exhibits significant renormalization of higher conduction bands, with CB+4 states either shifting outside the relevant energy window or becoming optically forbidden due to registry-dependent selection rules. This difference with the AB${_\text{Se}}$ stacking configuration highlights how variations in the atomic registry can dramatically modify the dimensionality of excitonic states. The absence of 3D excitons in AB${_\text{W}}$ clearly shows how stacking symmetry can be an effective way to control excitons in van der Waals heterostructures.

\section{CONCLUSION}
To summarize, we have systematically investigated the excitonic landscape in single-layer WSe$_{2}$/MoS$_{2}$ heterojunctions, focusing on the role of different stacking configurations. On the one hand, our analysis reveals that tightly bound excitons with an interlayer 3D character are uniquely present in the AA and AB${_\text{Se}}$ stackings, where the atomic registry enables strong out-of-plane hybridization via transition-metal d-orbitals (Mo/W) coupling. On the other hand, 2D excitons and interlayer charge-transfer excitons can be found in all stacking types, showing that in-plane quantum confinement and interlayer dipole coupling are strong regardless of differences in symmetry. These findings highlight the delicate interplay between structural arrangement and excitonic properties, providing key tools for engineering van der Waals heterostructures with tailored electronic and optical responses.

The flat and commensurate stacking configurations examined in the present study occur within real materials, as observed in transition metal dichalcogenides (TMDs) and their vdW heterostructures. \cite{Chendong2017,Yu2015,Alexeev2019,Alexeev2024} As a consequence, our results offer important insights into the electronic and optical properties of such microscopic regions, which can significantly influence the behavior of the overall system. This is particularly relevant in the context of Moiré crystals, which are distinguished by their complex interlayer arrangements and rotational alignments, as reported in TMD-based vdW heterojunctions. 
\cite{Chendong2017,Alexeev2017,Alexeev2019,Alexeev2024,Merkl2019,Yu2015,
	Deilmann2018,Gong2013,Kang2013,Yong2017,Chuu2017,Kozewa2016,Seyler2019,Tran2019,Jens2018,Pramoda2017,Heo2015,Fengcheng2015,Vanderzande2014}
Our results contribute to a deeper understanding of fundamental structure-property relationships at the cutting edge of ab initio many-body theory, building on recent advances in the study of TMD-based vdW heterostructures and their emergent phenomena. \cite{Merkl2019} Despite the fact that  these advanced computational methods are often challenging  for modeling superlattices with realistic unit cell sizes due to their substantial computational demands, they provide us with essential basic data for constructing model Hamiltonians and semiempirical approaches capable of addressing these complex systems, as highlighted in recent studies on TMD-based vdW heterostructures. \cite{Tscheppe2024,Forg2021,Crepel2024,Tawfiqur2022,Devakul2021,Ramos2025} These insights are crucial for advancing the design and understanding of next-generation quantum materials and devices based on TMD vdW heterostructures heterostructures.\cite{Withers2015}

\section{Acknowledgments}
This work has been been supported by the open research fund of the General Direction of Research and Technological Development (DGRSDT) of the Ministry of Higher Education and
Scientiﬁc Research, Algeria and by the Deutsche Forschungsgemeinschaft (DFG, German Research Foundation) through the research unit QUAST, FOR 5249 P7 ID No. 449872909. Ab initio simulations were performed using computational resources from the HPC facility at the University of Batna 2 (Algeria),  local clusters from CNR ISM at Monterotondo  (Italy), and the Head1 ITP cluster at Heidelberg University (Germany) under scholarship Coimbra 2024. W.L and K.R thank Dr Andrea Marini for his outstanding help  and valuable advice, W.L acknowledges Dr. Fulvio Paleari for technical assistance with the Yambo code and Yambopy toolkit .

\end{document}